\documentclass[11pt,a4paper,preprintnumbers,nofootinbib]{revtex4}
\pdfoutput=1

\usepackage[T1]{fontenc}
\usepackage[utf8]{inputenc}

\usepackage[english]{babel}

\usepackage{graphicx}
\usepackage[dvipsnames]{xcolor}
\usepackage{subfigure}

\usepackage{amsmath}
\usepackage{amsfonts}
\usepackage{amssymb}
\usepackage{amstext}

\usepackage{microtype}

\usepackage{multirow}
\usepackage{booktabs}

\usepackage{hyperref}
\usepackage{placeins}
\usepackage{verbatim}

\DeclareMathOperator{\Tr}{Tr}

\begin{document}

\title{Flavorful leptoquarks at hadron colliders}

\author{Gudrun Hiller}
\email{ghiller@physik.uni-dortmund.de}
\affiliation{Fakult\"at Physik, TU Dortmund, Otto-Hahn-Str.4, D-44221 Dortmund, Germany}
\author{Dennis Loose}
\email{dennis.loose@udo.edu}
\affiliation{Fakult\"at Physik, TU Dortmund, Otto-Hahn-Str.4, D-44221 Dortmund, Germany}
\author{Ivan Ni\v{s}and\v{z}i\'c}
\email{ivan.nisandzic@kit.edu}
\affiliation{ Institut  f\"ur Theoretische Teilchenphysik, Karlsruher Institut  f\"ur Technologie, D-76128 Karlsruhe, Germany}
\preprint{DO-TH 17/27}
\preprint{TTP18-008}

\begin{abstract}
$B$-physics data and flavor symmetries suggest that leptoquarks can have masses as low as few ${\cal{O}}(\mbox{TeV})$, predominantly decay to third generation quarks,
and highlight $pp \to b \mu \mu$  
signatures  from single production and $pp \to b b  \mu \mu$ from  pair production. Abandoning flavor symmetries could allow for inverted quark hierarchies, and cause sizable $pp \to j \mu \mu$ and  $jj \mu \mu $  cross sections, induced by second generation couplings. 
Final states with leptons other than muons including lepton flavor violation (LFV)  ones can also arise. The corresponding couplings can also be probed 
by precision studies of the $B \to (X_s, K^*,\phi) ee $  distribution and LFV searches in $B$-decays.
We demonstrate sensitivity in single leptoquark production for the LHC and extrapolate to the high luminosity HL-LHC.
Exploration of the bulk of the phase space requires a hadron collider beyond the reach of the LHC, with $b$-identification capabilities.

\end{abstract}

\maketitle

\section{Introduction}

Leptoquarks generically couple differently to different  generations of quarks and leptons.
Present hints of non-universality between electrons and muons in the rare $B$-decay observables $R_K, R_{K^*}$  \cite{Hiller:2003js}  by the LHCb collaboration  \cite{Aaij:2014ora,Aaij:2017vbb}  are indeed
naturally explained by  tree-level exchange of  leptoquarks \cite{Hiller:2014yaa, Gripaios:2014tna, Hiller:2016kry,Barbieri:2015yvd,Varzielas:2015iva,Dorsner:2016wpm,Fajfer:2015ycq,Becirevic:2016oho,Alonso:2015sja,Calibbi:2015kma,Sahoo:2015wya,Becirevic:2015asa,Cox:2016epl}.
Combining $R_K$ with  $R_{K^*}$ allows to diagnose the chirality of the participating $|\Delta b|=|\Delta s|=1$ currents \cite{Hiller:2014ula}. Current data favor
leptoquarks that couple to quark- and lepton doublets, {\it e.g.},\cite{Altmannshofer:2017yso,DAmico:2017mtc,Capdevila:2017bsm,Hiller:2017bzc,Geng:2017svp,Ciuchini:2017mik}
implying couplings to both $b$-  and $t$-quarks,  and charged leptons and  neutrinos.
The corresponding leptoquark representations are the scalar $SU(2)_L$-triplet $S_3$ and two vectors, a singlet $V_1$ and a triplet, $V_3$.
Importantly,  the mass scale of the leptoquarks is  model-independently limited by multi-$O(10)$ TeV, and in viable flavor models
in the TeV-range, suggesting dedicated searches at the Large Hadron Collider (LHC) and beyond \cite{Shiltsev:2017tjx}.

Search results  for  leptoquarks are available   for fixed branching fractions  into a given lepton species.
For instance, mass limits for leptoquarks  decaying  50 \%  to electrons (muons) plus jet  and the other  50 \% to neutrinos and a jet  are
900 GeV (850 GeV) \cite{Aad:2015caa}, with similar results reported in  \cite{Aaboud:2016qeg,Khachatryan:2015vaa}, all obtained from pair production.
The limits improve to 1100 GeV   \cite{Aaboud:2016qeg} (1080 GeV \cite{Khachatryan:2015vaa}) for 100 \% decay to electrons (muons) plus jet.
Bounds from single production in jet $ee$ (1755 GeV) and jet $\mu \mu$ (660 GeV) \cite{Khachatryan:2015qda} assume 100 \%  decays to charged leptons and
coupling equal to one.
As we will show, rare $B$-decay data suggest to look for leptoquarks with dominant decays to $b \ell$, $\ell=e,\mu$. To date, no corresponding  leptoquark search has been performed at the
LHC. On the other hand, derived limits from other searches such as supersymmetry resulting in analogous signatures as the leptoquark ones are 
1.5 TeV ($S_3$) and 1.8 TeV ($V_{1,3}$) for $b e$, and $1.4$ TeV  ($S_3$) and 1.7 TeV ($V_{1,3}$)  for $b \mu$  \cite{Diaz:2017lit}.
Limits on $b \nu, t \ell, t \nu$ are not stronger.

The aim of this study is to  work out  collider signatures of  leptoquark scenarios that  take into account flavor structure and $B$-physics data. We focus on single production, which is directly driven by the leptoquark couplings to quarks and leptons and results in signatures with  a quark and two leptons. Flavor physics provides
directions to identify  the  final states  with  leading signatures.
We work out explicit predictions for the scalar leptoquark $S_3$; the flavor aspects of the analysis are analogous for the vector ones.
We estimate improvements in mass reach for possible future $pp$-machines operating at center-of-mass energies $\sqrt{s}=33$ TeV (HE-LHC) and 100 TeV (FCC-hh) \cite{Shiltsev:2017tjx,Mangano:2017tke}.  For related recent works on leptoquark production and $R_{K,K^*}$, see \cite{Dey:2017ede,Buttazzo:2017ixm,Allanach:2017bta}.

The paper is organized as follows:
In section \ref{sec:S3} we review the requirements and constraints from flavor physics on the leptoquark's mass and couplings.
Leptoquark branching ratios and single as well as pair production are discussed in section \ref{sec:collider}.
Expectations for the flavor patterns of the leptoquark couplings are given in section \ref{sec:bench}, together with corresponding branching ratios and signal strengths.
In section \ref{sec:con} we conclude.

\section{The scalar leptoquark $S_3$ \label{sec:S3}}

We denote by $S_3$ the scalar leptoquark that resides in the $(\bar{3},3,1/3)$ representation of the SM gauge group. Its couplings to the SM fermions are given by the following Lagrangian:
\begin{equation}
	\mathcal{L}_{\text{Yuk}}=\lambda\,\bar{Q}^{C\,\alpha}_L (i \sigma^2)^{\alpha\beta}(S_3)^{\beta\gamma}L_L^\gamma+ Y_\kappa\,\bar{Q}^{C\,\alpha}_L (i \sigma^2)^{\alpha\beta}(S_3^\dagger)^{\beta\gamma}Q_L^\gamma+\text{h.c.},\label{S3-QL-Lagrangian}
\end{equation}
where $\sigma^2$ is  the second Pauli matrix and $\alpha,\beta,\gamma$ are $SU(2)_L$ indices, while $\psi^C$ denotes the charge conjugated spinor. We concentrate in this work on the first term that involves the coupling to leptons and quarks and assume the existence of a mechanism that forbids the second term that is potentially dangerous with regards to  proton decay. Our interest is therefore focused on the Yukawa coupling matrix $\lambda$, a  3 $\times $ 3  matrix in flavor space with rows (columns) carrying a quark (lepton) flavor index,
that we suppress for the moment to avoid clutter.
The $S_3$ can be represented in terms of its isospin components as
\begin{equation}
	S_3=
	\begin{pmatrix}
		S_3^{1/3} & \sqrt2 S_3^{4/3} \\
		\sqrt2 S_3^{-2/3} & -S_3^{1/3}
	\end{pmatrix} \, , 
\end{equation}
where the superscripts denote the electric charge in units of $e$. 
The normalization is fixed to yield  canonically normalized kinetic terms for the complex scalar components.

Expanding the Lagrangian~\eqref{S3-QL-Lagrangian} in terms of the isospin components we obtain
\begin{equation}
	\begin{split}
		\mathcal{L}_{\text{QL}}=-\sqrt2\lambda\,\bar{d}^C_L \ell_L\,S_3^{4/3} - \lambda\,\bar{d}^C_L\,\nu_L\, S_3^{1/3}
		+\sqrt2\lambda\,\bar{u}^C_L\,\nu_L\,S_3^{-2/3}-\lambda\,\bar{u}^C_L\,\ell_L\,S_3^{1/3}+\text{h.c.}
	\end{split}
	\label{S3-QL-Lagrangian-components}
\end{equation}

The kinetic term for the leptoquark multiplet is written as
\begin{equation}
	\mathcal L_\mathrm{kin} = \frac12\Tr\left[\left(D_\mu S_3\right)^\dagger D^\mu S_3\right]  \, . 
	\label{eq:S3_kin}
\end{equation}
We  assume the approximate mass degeneracy of the components within the multiplet.
For the collider study in section~\ref{sec:collider}
we implement the model (\ref{S3-QL-Lagrangian-components}), (\ref{eq:S3_kin}) in $\mathtt{Feynrules}$~\cite{Alloul:2013bka} to obtain the corresponding Universal Feynrules Output ($\mathtt{UFO}$) \cite{Degrande:2011ua}. The latter is used as input to
the $\mathtt{MadGraph}$ event generator code~\cite{Alwall:2014hca}. 

To successfully  accommodate present $R_{K^{(\ast)}}$ data with the $S_3$ one requires \cite{Hiller:2017bzc}
\begin{equation}
\lambda_{b\mu}\lambda_{s\mu}^\ast-\lambda_{be}\lambda_{se}^\ast \simeq 1.1 \frac{M_{S_3}^2}{(35\,\text{TeV})^2} \, .     \label{eq:RK_S3}
\end{equation} 
Here, we label 
the element of the leptoquark Yukawa matrix $\lambda=\lambda_{q \ell}$ by the quark and lepton flavors it couples to.
By $SU(2)_L$,   $\lambda_{U_i \ell} = V_{ji}^* \lambda_{D_j \ell}$, where $V$ denotes the CKM matrix, and  $U=u,c,t$, $D=d,s,b$ and $i,j=1,2,3$.
Assuming {\it i)} that the SM hierarchies for the quark Yukawas are intact in the leptoquark ones, couplings to third generation quarks are dominant
\cite{Varzielas:2015iva,Chankowski:2005qp}, 
\begin{align} \label{eq:suppression}
\lambda_{d \ell} \sim (\epsilon^3  \ldots \, \epsilon^4 ) \,  \lambda_{b \ell} \, , \quad \quad 
\lambda_{s \ell} \sim \epsilon^2 \,  \lambda_{b \ell} \, , ~~\ell=e, \mu , \tau \, .
\end{align}
This can, for instance, be realized with a Froggatt-Nielsen-Mechanism \cite{Froggatt:1978nt}, where $\epsilon\sim 0.2$ denotes a flavor parameter of the size of the sine of the Cabibbo angle. The $\sim$ symbol indicates that a relation holds up to factors of order one.
Charged lepton mass hierarchies are taken care of  by the  $SU(2)_L$-singlet leptons, {\it i.e.}, the lepton doublets are neutral under the
Froggatt-Nielsen symmetry and no further suppressions in $\lambda_{q \ell}$ appear.
Taking in addition into account that
{\it ii)}  the BSM effects in $R_{K,K^*}$ are predominantly from muons  as opposed to electrons as corresponding contributions are consistent with 
those from global fits to the $b\to s\mu^+ \mu^-$ observables \cite{global-fits}, a viable "simplified" benchmark $\lambda_s$ is obtained as 
\begin{equation}
	\lambda_s \sim  \lambda_0 \begin{pmatrix} 0 & 0 & 0 \\ * & \epsilon^2 & * \\ * & 1 & * \end{pmatrix}\,.
	\label{eq:coupling_pattern}
\end{equation}
Here the entries denoted by "0" are of higher order in $\epsilon$; they are constrained by $\mu$-$e$ conversion and rare kaon decays and of no concern to the  present analysis.
 The entries  labeled with an asterisk are not needed to explain $| \Delta b|=|\Delta s|=1$ data.
Eq.~(\ref{eq:RK_S3}) implies $\lambda_0\simeq M_{S_3}/ 6.7\,\text{TeV}$.
Allowing for order one factors in $\lambda_{s\mu}$, taken here to be between $1/3$ and $3$, one obtains the range
\begin{align} \label{eq:l0range}
M_{S_3}/ 11.6\,\text{TeV} \lesssim \lambda_0 \lesssim  M_{S_3}/ 3.9\,\text{TeV} \, .
\end{align}
The parameter space (\ref{eq:l0range}) is well within the LHC-limits on Drell-Yan production, to which $t$-channel leptoquarks contribute at tree level.
Specifically,  the Wilson coefficient $C_{b_LL} =v^2 \lambda_0^2/(2M_{S_3}^2)$  satisfies in our case  $C_{b_LL} \lesssim 2 \cdot 10^{-3}$, where $v=246$ GeV denotes the vacuum expectation value  (vev) of the Higgs,  while  experimentally it is constrained  only at the level of $10^{-2}$ for both electrons and muons  \cite{Greljo:2017vvb}. 
Note that the effective theory is  constructed to hold for leptoquark masses greater than the dilepton invariant mass, presently  up to a  few TeV. However, also for smaller masses
effective theory bounds provide  a useful approximation \cite{Bessaa:2014jya}.

\section{Collider signatures \label{sec:collider}}

We discuss leptoquark decays and single leptoquark production at proton-proton colliders in section \ref{sec:decay} and \ref{sec:single}, respectively.
In section \ref{sec:jets} we consider signatures with tops and jets.
We occasionally use the symbol  $\phi$ for a generic leptoquark.

\subsection{Decay and width \label{sec:decay}}

\begin{figure}[t]
\begin{center}
\vskip-.5cm
\includegraphics{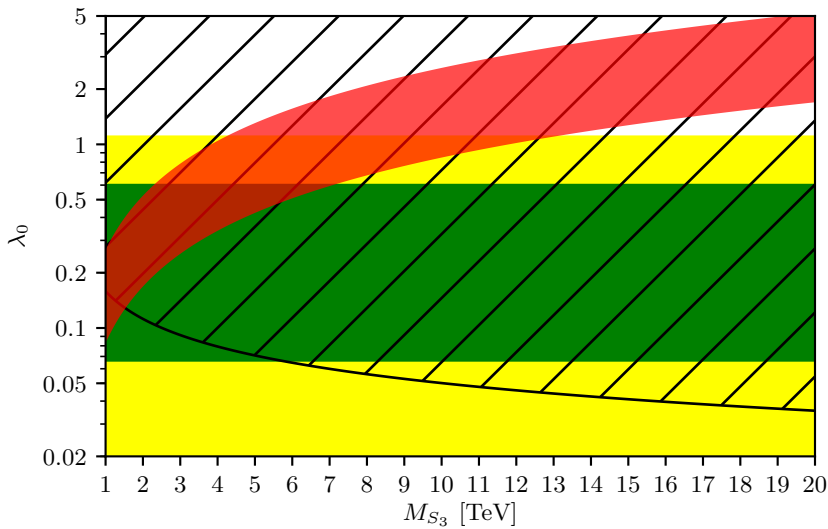}
\end{center}
\vskip-.8cm
\caption{Leptoquark parameter space, mass $M_{S_3}$  versus coupling,  in the simplified scenario assuming a single dominant coupling $\lambda_0$ for $S_3^{4/3}$.
The red band shows the range relevant to $R_K,R_{K^*}$-data, (\ref{eq:l0range}),  the  yellow (shaded) region refers to a narrow width $\Gamma/M_{S_3} \lesssim 5 \%$ , and in the hatched area above the black curve holds $\Gamma > \Lambda_{QCD}$. Flavor model predictions \cite{Hiller:2016kry} are given by the  green horizontal band. See text for details. }
\label{fig:space}
\end{figure}

Neglecting the masses of the decay products, the partial decay width of a scalar leptoquark $S_3$ with mass $M$ decaying  to a lepton $\ell$ and a quark $q$ reads
\begin{equation}
	\Gamma(S_3\to q\ell) =  c \frac{|\lambda_{q\ell}|^2}{16\pi}M\, ,
	\label{eq:LQ_decay_width}
\end{equation}
where $c=2$ for $S_3^{4/3}$, $S_3^{-2/3}$ and  $c=1$ for $S_3^{1/3}$, see eq.\ \eqref{S3-QL-Lagrangian-components}.
$\Gamma$ approximates the total width if the coupling $\lambda_{q \ell}$ is the dominant one. Note that the multi-body decays induced by inter-multiplet cascades such as
$S_3^{-4/3} \to S_3^{-1/3} W^- \to b \nu W^-$ can become sizable for large masses.
With  couplings to the first and second quark generation being flavor-suppressed, as, for instance, made explicit in (\ref{eq:coupling_pattern}) and following text, the leptoquark  predominantly decays to third generation quarks.
The relevant parameter space in mass  and leading coupling $\lambda_0 \equiv \lambda_{b \mu}$ is illustrated in figure \ref{fig:space}  for the $S_3^{4/3}$.
The yellow (shaded) region corresponds to a narrow width, $\Gamma/M_{S_3} \lesssim 5 \%$, which translates to $\lambda_0 \lesssim 1.1$.
(Note,  $\Gamma/M_{S_3} \lesssim 1 \% \, (10 \%)$ corresponds to $\lambda_0  \lesssim  0.5  \, (1.6)$.)
The red band denotes the region that explains lepton non-universality (LNU)-data, (\ref{eq:l0range}). In the hatched region above the black curve the leptoquark decays too rapid to form bound states,
$\Gamma > \Lambda_{QCD}$. Predictions from viable flavor models  $\lambda_{b \mu} \sim c_\ell$, where $c_\ell$ is of the order $\epsilon$  \cite{Hiller:2016kry} (green horizontal band) are also shown. 
$B_s$-mixing data together with $R_K,R_{K^*}$ provide a data-driven upper limit on the mass of the $S_3$ leptoquark of 40 TeV.
For such large masses the coupling required by $B$-physics data becomes order one and approaches the perturbativity limit.
In addition, the region of narrow width is left.
Upper mass limits on the (gauge-like) vector leptoquarks are 45 TeV and 20 TeV for $V_1$ and $V_3$, respectively  \cite{Hiller:2017bzc}~\footnote{Recent analysis of
$B_s$-mixing \cite{DiLuzio:2017fdq} suggests lower upper limits on the leptoquark masses.}.

We list the dominant decays modes for the three leptoquark representations that can explain current LNU-data  \cite{Hiller:2017bzc},
for the scalar isospin triplet
 \begin{eqnarray}
S_3^{+2/3} &\to& t\ \nu, \nonumber \\
S_3^{-1/3} &\to& b\ \nu \ , \  t\ \mu^- ,     \label{eq:S3decay}\\
S_3^{-4/3} &\to& b\ \mu^-, \nonumber 
\end{eqnarray}
the vector isospin singlet
\begin{eqnarray}
V_1^{+2/3} &\to& b\ \mu^+ \ , \  t\ \nu    \label{eq:V1decay}
\end{eqnarray}
and the vector isospin triplet
\begin{eqnarray}
V_3^{-1/3}   &\to& b\ \nu \nonumber, \\
V_3^{+2/3}   &\to&   b\ \mu^+ \ , \  t\ \nu,    \label{eq:V3decay}\\
V_3^{+5/3}   &\to& t \mu^+.\nonumber
\end{eqnarray}
As, for instance, $V_1^{-2/3} \to \bar b\ \mu^- $ and  $S_3^{-4/3} \to b\ \mu^- $ lead both to a negatively charged lepton, tagging of the $b$-charge would be useful to identify the leptoquark type and its electric charge.

Note that some leptoquarks can undergo more than one decay into the third generations quarks, such as
$S_3^{-1/3}$, (\ref{eq:S3decay}) and in this case, by $SU(2)_L$,  $\mathcal B(\phi \to b \nu_\ell) \sim \mathcal B(\phi \to t \ell)  \simeq 1/2$.
Similarly, for $V_1$  (\ref{eq:V1decay}) and $V_3$  (\ref{eq:V3decay}), $\mathcal B(\phi \to b  \ell) \sim \mathcal B(\phi \to t  \nu_\ell)  \simeq 1/2$.

\subsection{Single leptoquark  production \label{sec:single}}

\begin{figure}[h]
	\begin{center}
		\subfigure[t][]{\includegraphics[width=0.25\textwidth]{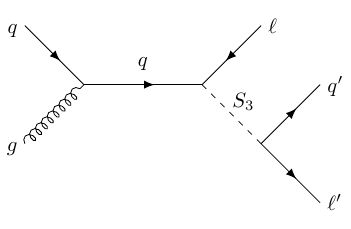}}
		\hspace{.6cm}
		\subfigure[t][]{\includegraphics[width=0.25\textwidth]{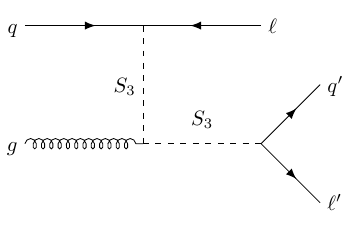}}
		\hspace{.6cm}
		\subfigure[t][]{\includegraphics[width=0.25\textwidth]{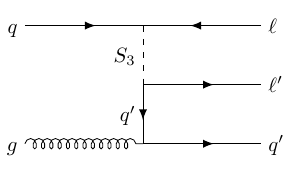}}
	\end{center}
	\caption{Leading order diagrams for single leptoquark production and decay: Diagrams a), b)  correspond to resonant amplitudes. 
	Diagram c) corresponds to a non-resonant contribution, the effects of which  are suppressed through  kinematic cuts, see section \ref{sec:single} for details.
	}
	\label{fig:leading}
\end{figure}

\begin{figure}[h]
	\begin{center}
	\includegraphics{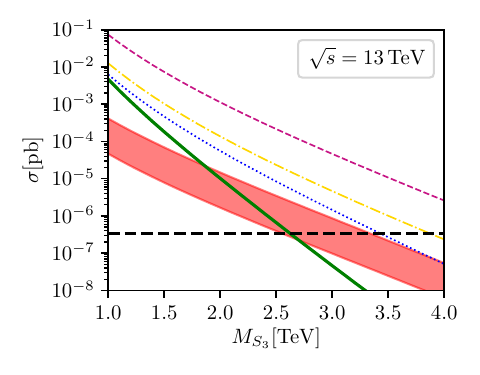}
	\includegraphics{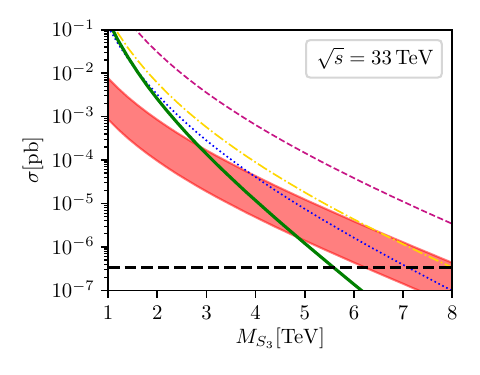}
	\includegraphics{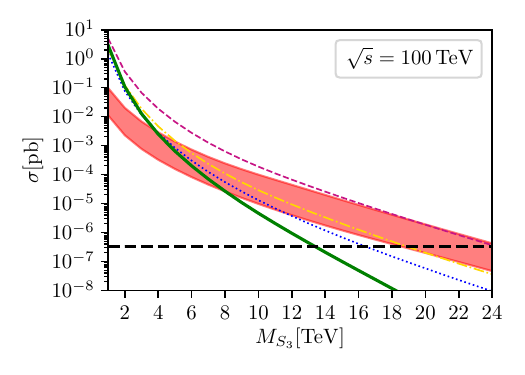}
	\end{center}
	\caption{The single leptoquark  production cross section $\sigma(pp\to S_3^{-4/3}\mu^+ + S_3^{+4/3}\mu^-)$ as a function of the mass $M_{S_3}$ 
	for $\sqrt{s}=13, 33$ and 100 TeV. The red band corresponds to  the flavor pattern  \eqref{eq:coupling_pattern} with  $\lambda_0$  in accord with the $B$-anomalies (\ref{eq:l0range}). 
	The triplet of (thin) curves illustrates the single production cross section with one coupling switched on at a time  (from top to bottom: dashed pink, dashed-dotted orange and dotted blue for $\lambda_{d\mu, s\mu, b\mu}$ set to one, respectively). The pair production cross section $\sigma(pp\to S_3^{-4/3} S_3^{+4/3})$ is shown by the  green (thick, solid) curve. 
	The black (dashed)  line corresponds to the absolute lower limit of  the cross section below which one  cannot produce a single event with integrated luminosity $3000\,\text{fb}^{-1}$. See text for details.}
		\label{fig:quark_gen_comparison}
\end{figure}

In figure \ref{fig:leading} we show the leading order diagrams inducing single leptoquark production, followed by its decay.
The production is in association with a lepton. The cross section is sensitive to the flavor coupling $\lambda_{q \ell}$.
With the couplings to the first and second quark generations being flavor-suppressed, the parton level production of the leptoquark is dominated by the third generation coupling.
This continues to be the case at hadron level, which
 can be inferred  from figure \ref{fig:quark_gen_comparison}. The parton distribution function (PDF) suppression of $b$-production (dotted blue curve) versus $s$-production (dashed-dotted orange curve)  is about a factor of $(\mbox{few})^{-1}$
and the one of  $b$-production versus $d$-production (dashed pink curve) is of the order $10^{-1}$ to $10^{-2}$, which are weaker than the respective flavor suppressions (\ref{eq:suppression}),
so indeed, beauty wins.  In figure \ref{fig:quark_gen_comparison} we added the cross sections of CP-conjugate final states; in the absence of CP-violation, which is  the limit we are working in, this amounts to a factor $2$ in the single production cross section from valence quark-gluon fusion.
Also shown in the two additional plots are predictions for future proton-proton machines, a 33 TeV HE-LHC and a 100 TeV collider.

The corresponding numerical calculations are performed using $\mathtt{Madgraph\,v.2.6}$ \cite{Alwall:2014hca} at leading order in QCD. 
We find that the largest uncertainties originate from the PDFs (we use $\mathtt{LHAPDF}$~\cite{Buckley:2014ana}). For the single production (red band)  linked to  $R_{K^{(*)}}$-data  (\ref{eq:l0range}) they grow from order ten percent for $M \sim 1$ TeV to $\sim 35-40$ percent for smaller cross sections  of  $\mbox{few} \times 10^{-7} \, {\rm pb}$.
The scale uncertainty --  in our estimate both the factorization and the  renormalization scale  are equal  to  half of the sum of the transverse masses of the final state particles --
reaches $\sim 25$ percent.

 In figure \ref{fig:quark_gen_comparison} the cross section for pair production $\sigma(pp\to S_3^{-4/3} S_3^{+4/3})$ is shown by the  solid green curve.
  We find, using Madgraph at leading order, that both PDF and scale uncertainties can reach ${\cal{O}}(40)$ percent towards $\sigma \sim \mbox{few} \times 10^{-7} \, {\rm pb}$.
While the scale uncertainty is essentially flat the PDF uncertainty drops to order 10 percent for lighter leptoquarks near a TeV.
 In the  simplified benchmark (\ref{eq:coupling_pattern}) the $S_3^{ \pm 4/3}$ decays  into $b \mu$,  see (\ref{eq:S3decay}), producing a
 $pp \to bb \mu \mu$ signature.
  Pair production of another component of the $S_3$ can give $tt \mu \mu$, $bb E_{\rm miss}$, $b  t \mu E_{\rm miss}$ or $tt E_{\rm miss}$ final states.
 
 For low masses, pair production  has a larger cross section than single production (red band)  linked to  $R_{K^{(*)}}$-data  (\ref{eq:l0range}), while the single production
 cross section is larger for  higher masses.
Naively, about  a factor $\sim 2$ ($5$) in mass reach can be gained  in pair production 
at a  $33\,(100)$ TeV collider relative to $13$ TeV and for comparable luminosity of $3000 \,  \mbox{fb}^{-1}$. The potential gain for single production is somewhat larger:
about  a factor $\sim 2.5$ ($7$)  in the target parameter space  - the red band -
for $33\,(100)$ TeV. While this gives an idea about the accessible ranges dedicated simulations are needed to estimate the reach more reliably.

\begin{figure}[h]
	\begin{center}
		\subfigure[t][]{\includegraphics[width=0.43\textwidth]{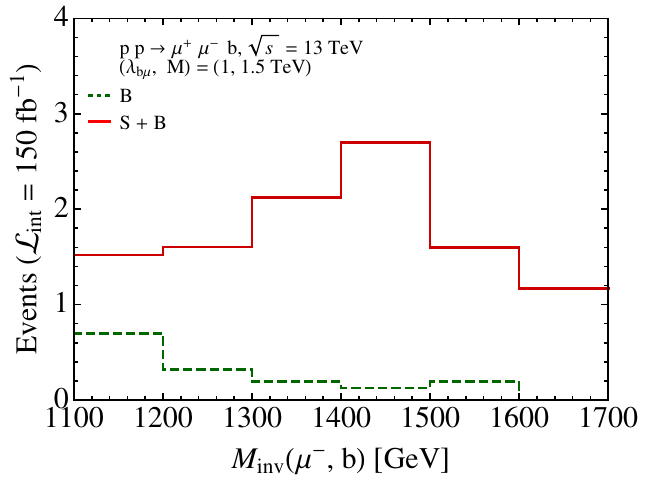}}
		\hspace{.25cm}
		\subfigure[t][]{\includegraphics[width=0.43\textwidth]{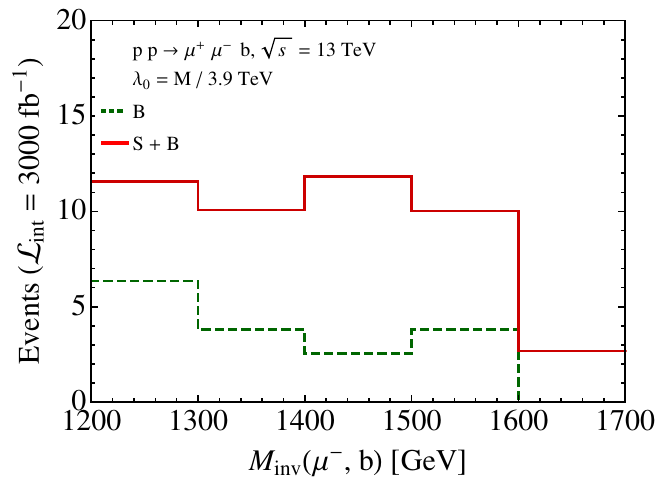}}
		\hspace{.25cm}
		\subfigure[t][]{\includegraphics[width=0.43\textwidth]{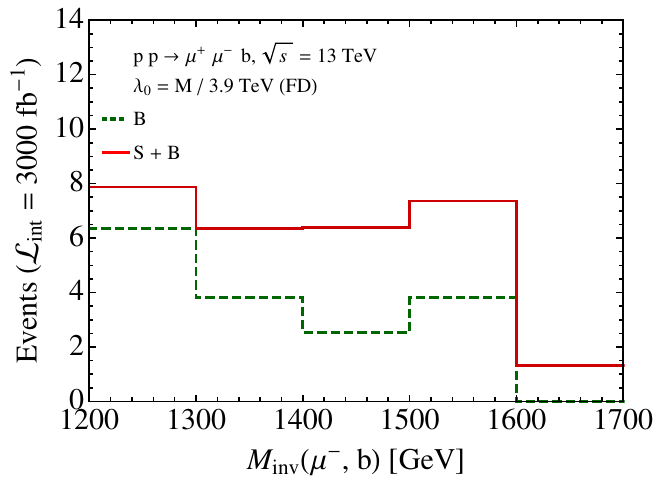}}
	\end{center}
	\caption{Distribution of events over the invariant mass of the $\mu^-  \text{-}b\text{-jet}$ system from $p p \to b \mu^+\mu^-$ and $p p \to \bar{b} \mu^+\mu^-$ at  $\sqrt{s}=13\,\text{TeV}$. All three plots correspond to a leptoquark with mass $1.5\,\text{TeV}$. a) corresponds to $\lambda_{b\mu}=1$, and  b) to the  flavor pattern \eqref{eq:coupling_pattern} with $\lambda_0=M/3.9$ TeV in accord with $B$-anomalies (\ref{eq:l0range}).  The pattern  \eqref{eq:hierarchy} on which  c) 
	is based similarly addresses $B$-data but allows additionally for decays to taus. The dashed green (solid red) line represents background (signal and background) events.
	Kinematic cuts are described in the text. 
	}
	\label{fig:distributions-mub_2}
\end{figure}

We simulate events for a $1.5$ TeV leptoquark and different couplings  at the $\sqrt{s}=13$ TeV LHC.
In figure \ref{fig:distributions-mub_2} we present the corresponding distributions for  signal and  background as a function of $M_{\text{inv}}(\mu^-,b)$, the invariant mass of the 
$\mu^- \text{-}b-$system. To enhance the significance and to study a situation where the $b$-charge is not tagged, we add the CP-conjugate process $pp \to \bar b \mu^+ \mu^-$ in both signal and background. The corresponding calculations are performed at leading order in QCD using $\mathtt{Madgraph~v.2.6}$~\cite{Alwall:2014hca} for the event generation, $\mathtt{PYTHIA~8}$~\cite{Sjostrand:2014zea} for the parton showering and hadronization and $\mathtt{DELPHES~3}$~\cite{deFavereau:2013fsa} for the fast detector simulation.  For the muon isolation we 
follow~\cite{Khachatryan:2015qda}, while all the other criteria are taken from the default Delphes card for the CMS detector. To account for the QCD corrections at NLO we multiply signals and backgrounds with corresponding global k-factors. For the signal we use $k \sim1.5$ \cite{Hammett:2015sea}, while for the background we use the value that we obtain from comparing the NLO and LO calculations at fixed order in Madgraph. For the analysis of the signal and background we use $\mathtt{MadAnalysis~5}$~\cite{Conte:2012fm}. 

For these evaluations we adopt the following kinematic cuts: We accept events containing two opposite charge muons and a $b$-jet and require for the transverse momenta and absolute pseudo-rapidities of each of these final states to exceed $45\,\text{GeV}$ and to be smaller than $2.1$, respectively. Furthermore, for the angular separation between muon and a $b$-jet we have $\Delta R > 0.3$ and also we cut on the invariant mass of the opposite-charge muon-pair $M_{\text{inv}} > 110\, \text{GeV}$. For the final event selection we adopt the cut on the scalar sum of the transverse momenta of muon-pair and leading $b$-jet $S_T(\mu_1, \mu_2, b_1) > 250\,\text{GeV}$~\cite{Khachatryan:2015qda}. We use $M_{\text{inv}}(\mu, b\text{-jet})$ as a discriminating variable between signal and background and adjust the corresponding cut to a given mass and coupling hypothesis in order to maximize the significance of the signal.

The approximate expected discovery significance for the first case (a) $\lambda_{b \mu}=1$, which is a priori unrelated to LNU-data, is around $4\,\sigma$ for an integrated luminosity of $150\,\text{fb}^{-1}$.
In the second case (b) where $\lambda_{b\mu}$ saturates the upper limit in  \eqref{eq:coupling_pattern},(\ref{eq:l0range}) we find around $5\,\sigma$ at $3000\,\text{fb}^{-1}$. The significance for the third case c) is smaller, somewhat below $3\,\sigma$ due to the twice smaller branching fraction into $b\mu$, see \eqref{eq:hierarchy}.
To compute these significances we used the approximate formula from \cite{Cowan:2010js} and took into account both $\mu^- \text{-}b-$ and $\mu^+ \text{-}\bar b-$ signals
in the data sample.

\subsection{Tops and jets \label{sec:jets}}

We briefly discuss single production into  top and  jet plus dilepton final states, which complements searches into $b$'s.

The processes $g b \to \mu^+ \phi (\to b \mu^-)$ are related to
$g b \to \nu \phi (\to t \mu^-)$ by $SU(2)_L$, and arise at the same order of flavor counting. We recall the factor $1/2$ in the leading $S_3^{-1/3}$ branching ratios as within our approximations
the leptoquark  decays to both $b \nu$ and $t \ell$ via $\lambda_0$ at equal rate.
However, the  $t \mu \nu$ final state has larger background because SM processes  from $W \to \ell \nu$ cannot be removed equally well as $Z \to \ell \ell$.
On the other hand, if the flavor suppression of the second generation quark coupling is not realized in nature, $t \mu \mu$ final states, induced by 
$\lambda_{c \mu} \simeq \lambda_{s \mu}$,
could potentially be  interesting.  
For $\lambda_{b \mu}=\lambda_{s \mu}$  one finds for the branching ratio ${\cal{B}}(\phi \to t \mu) \simeq 1/4$, however, the  leptoquark coupling drops by an order of magnitude, 
$\lambda_0=0.03 M/\mbox{TeV}$, as dictated by  \eqref{eq:RK_S3}. Despite the PDF enhancement from charm relative to $b$ this democratic scenario
results in about two orders of magnitude smaller cross sections relative to $pp \to b \mu \mu$.

Inverted hierarchies $\lambda_{s \mu} \gg \lambda_{b \mu}$  are in conflict with flavor symmetry, see section \ref{sec:bench}, but  not excluded experimentally in the simplified scenario with two entries only (\ref{eq:coupling_pattern}). This extreme scenario however does not improve the situation regarding $t \mu \mu$, as the
branching ratio into $t \mu$ is suppressed as $|\lambda_{b \mu}/\lambda_{s \mu}|^2 $ while the product of couplings is fixed by $B$-data (\ref{eq:RK_S3}).
On the other hand, jet plus dileptons benefits from the large second generation Yukawa while having an order one branching ratio.
Note that jet plus charged lepton final states can arise from several components of the $SU(2)_L$-multiplets (\ref{eq:S3decay})-(\ref{eq:V3decay}). Using the  upper limit $\lambda_{s \mu} \lesssim M/2$ TeV   from Drell-Yan production at the LHC \cite{Greljo:2017vvb} and taking into account the PDF enhancement
 second quark generation  cross sections $\sigma(pp \to (j \mu^-  )\mu^++ (j \mu^+) \mu^-)$ can be about an order of magnitude larger than  the maximum
 third generation ones in accord with $B$-anomalies  (\ref{eq:l0range}), shown by the 
 red band in figure \ref{fig:quark_gen_comparison}.
A detailed analysis of the sensitivity to inverted hierarchies including reconstruction efficiencies  is left for future work.

\FloatBarrier

\section{Flavor benchmarks \label{sec:bench}}

We explain how  the simplified pattern (\ref{eq:coupling_pattern}) can arise in models of flavor and give more general Yukawa patterns.
We are in particular interested in theoretical predictions for the  entries with an asterisk, that potentially induce leptoquark signals with electrons or taus, and LFV, which affects the collider phenomenology.

The most general approach, treating all entries as free parameters only  constrained by upper limits and (\ref{eq:RK_S3}), presently does not suffice to identify
the dominant collider signatures. We therefore suggest to study benchmarks. They are  motivated by  viable flavor symmetries that successfully
explain SM flavor, and  consistency with data.

A simultaneous explanation of the LNU-ratios $R_K,R_{K^*}$ and  $B \to K^* \mu^+ \mu^-$ angular distributions is possible with
BSM effects in couplings to muons alone. Hence, from this perspective, leptoquark couplings to electrons are {\it not necessary}.
Lepton species isolation patterns can be engineered with discrete, non-abelian  flavor symmetries such as $A_4$ \cite{Varzielas:2015iva}.
For second generation leptons, these read
 \begin{eqnarray} \label{eq:single}
\lambda_\mu \sim c_\ell \left( 
\begin{array}{ccc}
0 &  \epsilon^4  & 0\\
0 &  \epsilon^2  & 0\\
0  &  1 & 0
\end{array}
 \right)  \,  \rightarrow  
 c_\ell \left( 
\begin{array}{ccc}
\delta   \epsilon^4 &  \epsilon^4  & \delta   \epsilon^4\\
\delta   \epsilon^2  &  \epsilon^2  & \delta   \epsilon^2 \\
\delta   &  1 & \delta 
\end{array}
 \right) 
\end{eqnarray}
or,  one that  avoids the CKM suppression for the second generation quarks \cite{Hiller:2016kry},
\begin{align} \label{eq:tildeLmu}
    \tilde \lambda_\mu &\sim   \begin{pmatrix} 0 & c_\ell \epsilon^4 & 0 \\ c_\nu\kappa & c_\nu\kappa & c_\nu\kappa  \\ 0 &  c_\ell  &0\end{pmatrix}   \quad
    \rightarrow 
    \begin{pmatrix}c_{\nu} \kappa \epsilon^{2} & c_{\ell} \epsilon^{4} + c_{\nu} \kappa \epsilon^{2} & c_{\nu} \kappa \epsilon^{2}\\ c_{\nu} \kappa & c_{\ell} \epsilon^{2} + c_{\nu} \kappa & c_{\nu} \kappa\\ c_{\ell} \delta + c_{\nu} \kappa \epsilon^{2} & c_{\ell} & c_{\ell} \delta + c_{\nu} \kappa \epsilon^{2}\end{pmatrix}  \, .
\end{align}
In (\ref{eq:single}) and (\ref{eq:tildeLmu})  all  vevs $c_\ell, c_\nu,\kappa$ are of the order $\epsilon^n$, $n \geq 1$. 
$\delta$ is a small parameter of  second order in the vevs, see  \cite{Hiller:2016kry} for details.

Patterns (\ref{eq:single}) and (\ref{eq:tildeLmu}) receive corrections from rotating  flavor to mass basis and from higher order spurion insertions  \cite{Hiller:2016kry},
both of which are incorporated  in the matrices to the right of the arrow.
As a result, in addition to those to muons, leptoquarks couple to all leptons and third generation quarks, and  LFV arises.

For the third benchmark we employ a more general parametrization and impose experimental constraints \cite{Varzielas:2015iva} "flavor data"
\begin{eqnarray}  \label{eq:hierarchy}
 \lambda_{\rm FD}=\lambda_0\left( 
\begin{array}{ccc}
\rho_d \kappa_e &  \rho_d   & \rho_d  \kappa_\tau \\
\rho \kappa_e &  \rho   & \rho \kappa_\tau  \\
\kappa_e &  1 & \kappa_\tau
\end{array} 
\right)  \, , \quad \quad \kappa_\tau\sim 1 \, .
\end{eqnarray}
Here we allow for quark flavor suppressions $\rho_d =\lambda_{d \ell}/\lambda_{b \ell}$ and $\rho =\lambda_{s \ell}/\lambda_{b \ell}$,  with 
 larger couplings for higher generations, in concordance with the observed quark mass pattern. In addition a suppression factor $\kappa_e$ for the electrons is accounted for.
 The phenomenologically viable range for $\lambda_{\rm FD}$ parameters in Eq.~(\ref{eq:hierarchy}) is \cite{Varzielas:2015iva} 
 \begin{eqnarray} \label{eq:param}
 \rho_d \lesssim 0.02 \, , \quad \kappa_e  \lesssim  0.5 \, , \quad 10^{-4} \lesssim \rho \lesssim 1  \, , \quad \kappa_e/\rho \lesssim  0.5   \, , \quad \rho_d / \rho \lesssim  1.6 \, . 
\end{eqnarray}
The MEG experiment \cite{Baldini:2013ke} can  put a limit on $\kappa_e/\rho$ at the level of 0.2 in the future \cite{Varzielas:2015iva}.

\begin{table}
	\centering
	\begin{tabular}{l||c|c|c||c|c|c}
 & $b \mu$  & $b e $ & $b \tau$ & $j \mu $ & $j e$  &  $j \tau$ \\
		\midrule
		$ \lambda_\mu$& 1 & $\delta^2$ & $\delta^2$ & $\epsilon^4$& $\epsilon^4 \delta^2$& $\epsilon^4 \delta^2$  \\
					
				$ \tilde \lambda_\mu$& 1  & $\delta^2$ & $\delta^2$ & $(c_\nu \kappa / c_\ell)^2$& $(c_\nu \kappa / c_\ell)^2$& $(c_\nu \kappa / c_\ell)^2$ \\
				$ \lambda_{\rm FD}$&1/2 & $\kappa_e^2/2$& 1/2& $\rho^2/2$&  $\rho^2 \kappa_e^2/2$&$\rho^2/2$  \\
					\end{tabular}
	\caption{Branching fractions of leptoquark $S_3^{-4/3}$ decaying to  $b \ell$ and $j \ell$, $\ell=e,\mu,\tau$ 
	for different flavor benchmarks  (\ref{eq:single}),  (\ref{eq:tildeLmu}) and (\ref{eq:hierarchy}), see text for details.
	Corresponding branching fractions of the $S_3^{-1/3}$ satisfy ${\cal{B}}(S_3^{-1/3} \to t \ell), {\cal{B}}(S_3^{-1/3} \to b \nu)  \sim {\cal{B}}(S_3^{-4/3} \to b \ell)/2$
	and for jets ${\cal{B}}(S_3^{-1/3} \to j \ell),  {\cal{B}}(S_3^{-1/3} \to j \nu) \sim {\cal{B}}(S_3^{-4/3} \to j \ell)/2$.}
		\label{table:br}
\end{table} 

Branching fractions of $S_3$ to $b \ell$ and $j \ell$, $\ell=e,\mu,\tau$  in the three benchmarks are presented  in table \ref{table:br}. 
They are also
useful  to estimate signatures in leptoquark pair production. Predictions for the vector leptoquarks follow  analogous flavor patterns:
Modulo the slightly different  vevs \cite{Hiller:2016kry} in $\tilde \lambda_\mu$ one obtains in addition
	 ${\cal{B}}(V_1 \to b \ell) \sim {\cal{B}}(S_3^{-4/3} \to b \ell)/2$ for similar masses.
In table \ref{table:signal}  we give the parametric signal strength for single leptoquark production using the narrow-width approximation
\begin{equation}
	\sigma(pp\to\phi(\to q\ell)\ell) = \sigma(pp\to\phi\ell)\mathcal B(\phi\to q\ell)
	\label{eq:nwa_factorization}
\end{equation}
 in the  benchmarks for different final state flavors. 
In both   tables \ref{table:br} and   \ref{table:signal} we give the leading terms
in the 
vev expansion,  $c_\ell, c_\nu, \kappa <1$, and the flavor factors $\rho,\rho_d, \kappa_e <1$.

Hierarchies in $\lambda_\mu$ and $\tilde \lambda_\mu$ are identical  for all $b$-final states, but the jet 
\footnote{We use 'jet'  for an object from gluons, $u,d,s$ and $c$-quarks and anti-quarks, as opposed to  a $b$-jet, made out of $b$ and $\bar b$.} signals are less suppressed in $\tilde \lambda_\mu$.
$b ee$ and $j ee$ channels  are strongly suppressed  in both benchmarks (\ref{eq:single}) and (\ref{eq:tildeLmu}).
For  $\lambda_{\rm  FD}$ the situation depends on how strong  flavor suppressions are. 
A small $\rho$ implies a suppressed $\kappa_e$, (\ref{eq:param}).
We identify
two limits:\\
A) $\rho,\kappa_e$ are order one, then either $\lambda_0$ has to be small,  or, if  $\lambda_0$ is order one as well, then leptoquark  masses are in the multi-10 TeV range. In either case there 
is  no  leptoquark-induced $b \mu \mu$ signal   at the LHC.\\
B) $\rho,\kappa_e \ll 1$, then  $\lambda_0$ is sizable, while leptoquark masses can be TeV-ish, and the jet and electron modes are suppressed.

Case B) resembles the situation for benchmarks $\lambda_\mu$ and $\tilde \lambda_\mu$.
Constraints on $\kappa_e$ are therefore important to control final states with electrons. For $\kappa_e \ll1$ the $ee$ or $e \mu$ modes would be SM-like.
$\kappa_e$ can be constrained from $b \to s ee$ or $b \to s e \mu$ processes together with $b \to s \mu \mu$.
Due to reduced uncertainties angular observables in $B \to K^*(\to K \pi) ee$ decays are promising \cite{Hiller:2014ula}.

\begin{table}
	\centering
	\begin{tabular}{l||c|c|c|c|c|c||c|c|c|c|c|c}
 & $b \mu \mu$  & $b e \mu$ & $b \tau \mu$ & $b ee$ & $be \tau$ &  $b\tau \tau $ & $j \mu \mu$ & $j e \mu$ & $j  \tau \mu$  & $j ee$ & $j e \tau$ & $ j \tau \tau$\\
		\midrule
		$ \lambda_\mu$& $c_\ell^2$ &$c_\ell^2 \delta^2$ &$c_\ell^2 \delta^2$  & $c_\ell^2 \delta^4$ & $c_\ell^2 \delta^4$ & $c_\ell^2 \delta^4$ & $c_\ell^2 \epsilon^4$ &$c_\ell^2  \delta^2 \epsilon^4$ &$c_\ell^2  \delta^2 \epsilon^4$  & $c_\ell^2  \delta^4 \epsilon^4$ & $c_\ell^2  \delta^4 \epsilon^4$ & $c_\ell^2  \delta^4 \epsilon^4$\\  
			
				$ \tilde \lambda_\mu$& $c_\ell^2$ &$c_\ell^2 \delta^2$ &$c_\ell^2 \delta^2$ & $c_\ell^2 \delta^4$ & $c_\ell^2 \delta^4$ & $c_\ell^2 \delta^4$ & $(c_\nu \kappa)^2$ & $(c_\nu \kappa)^2$& $(c_\nu \kappa)^2$& $(c_\nu \kappa \delta)^2$ & $(c_\nu \kappa \delta)^2$ & $(c_\nu \kappa \delta)^2$\\            
				$\lambda_{\rm FD}$ & $\lambda_0^2/2$ & $\lambda_0^2 \kappa_e^2/2$ & $\lambda_0^2/2$ & $\lambda_0^2 \kappa_e^4/2$ & $\lambda_0^2 \kappa_e^2/2$  & $\lambda_0^2/2$ & $\lambda_0^2 \rho^2/2$  &  $\lambda_0^2 \rho^2 \kappa_e^2/2$  &  $\lambda_0^2 \rho^2/2$  &$\lambda_0^2 \rho^2 \kappa_e^4/2$   & $\lambda_0^2 \rho^2 \kappa_e^2/2$   & $\lambda_0^2 \rho^2/2$\\
	\end{tabular}
	\caption{
	Parametric signal strength of $pp  \to b \ell \ell^\prime$  and  $pp \to j \ell \ell^\prime$ final states from single leptoquark $S_3^{-4/3}$ production for different flavor benchmarks  (\ref{eq:single}),  (\ref{eq:tildeLmu}) and (\ref{eq:hierarchy}), see text for details.  }
		\label{table:signal}
\end{table} 

In the presence of sizable couplings to more than one lepton species, such as muons and taus  there are two main
aspects to single production:
Firstly, the signal in  $pp \to (\phi \to b \mu) \mu$ drops because the leptoquark  has also a decay rate into $b \tau$.
This happens in plot c) of  figure \ref{fig:distributions-mub_2}.
Secondly, LFV arises, such as $pp \to (\phi \to b \tau) \mu$ and $pp \to (\phi \to b \mu) \tau$.
This can be searched for in a complementary way in $B \to K^{(*)} \tau \mu$ decays.

Up to  cuts and detection efficiencies, $S_3$-induced
 $pp \to t \mu \mu$, which arises from charm quarks, see figure \ref{fig:leading},  is suppressed by $\epsilon^4$, $(c_\nu \kappa/ c_\ell)^2$  and $\rho^2$ in scenario $\lambda_\mu$, $\tilde \lambda_\mu$ and
$\lambda_{\rm fd}$, respectively, with respect to $pp \to b \mu \mu$. Using the same approximations, the $pp \to t \mu \nu$ and
$pp \to b \nu \nu$ signal strength is the same as  for $pp \to b \mu \mu$.

Both $S_3^{-4/3}$  and $S_3^{-1/3}$  produce $j \ell \ell^{(\prime)}$ final states (\ref{eq:S3decay}), hence the parametric signal strength of all jet modes in table \ref{table:signal} is additionally enhanced by a factor
$\sim 3/2$. Here we used that the strange and charm PDFs are  similar in size within our approximations.

\section{Conclusion \label{sec:con}}

TeV-mass leptoquarks can be singly produced at hadron colliders in association with a lepton. 
 $B$-physics data, which hint at a BSM contribution in $b \to s \mu \mu$ processes, while one in  $b \to s ee$
 may be discarded by Occam's razor, together with  flavor model building identify  $pp \to \phi \mu \to b \mu \mu$,  and  two modes with missing energy $pp \to \phi \nu \to b \nu \nu$
  and  $pp \to \phi \nu \to t \mu \nu$, as the  channels with leading cross sections.
 While this highlights $b \mu \mu$ as a prime channel,  signatures with further final states can also be sizable  and should be explored. The reasons are, firstly, to advance our understanding of  flavor   by probing lepton and quark flavor specific couplings in the  leptoquarks' Yukawa matrix, 
as opposed to rare decays (\ref{eq:RK_S3}), which constrain products of couplings.
Secondly, exploration of further single production modes serves as cross check with other measurements, such as leptoquark pair production and indirect searches, notably
Drell-Yan production and semileptonic rare $b$-decays.

Due to the higher cross section for lower masses, see  figure \ref{fig:quark_gen_comparison}, we encourage searches for leptoquarks from  pair production decaying to a $b$-quark
 and a  lepton, or a top quark and a lepton, as in  (\ref{eq:S3decay})-(\ref{eq:V3decay}).
 
 Semileptonic $b$-decays
can be probed at LHCb and Belle II, and allow to access lepton specific couplings of all three generations,
 which could improve benchmarks (\ref{eq:hierarchy}) and aid collider searches.
Corresponding processes are $b \to s ee$ and LFV, $b \to s e \mu$ and those into taus.
Studies of the angular distribution in $B \to K^* ee$ similar to $B \to K^*  \mu \mu $ \cite{Hiller:2014ula,Wehle:2016yoi} and searches for  $B \to K^{(*)} e (\mu, \tau)$ and
$B \to K^{(*)} \mu \tau $
at the level of $10^{-8}$ and lower,
and $B_s \to e \mu$ at  ${\cal{O}}(10^{-11})$ \cite{Varzielas:2015iva} should be pursued
to obtain meaningful constraints on the leptoquark flavor matrix. 

 Discrimination between different quark flavors can be achieved  by comparing signatures induced by the third  generation quark 
 coupling to  the ones induced by first two  generations, such as $b \ell \ell $  to  $j \ell \ell$, 
 respectively.  Flavor symmetries predict the latter to be suppressed or at least not enhanced relative to the former, as a result of  the corresponding quark hierarchies, {\it e.g.,}
 see patterns  (\ref{eq:single}),  (\ref{eq:tildeLmu}) and (\ref{eq:hierarchy}).
An experimental search could put this  prediction to a test. Evidence for an inverted quark hierarchy $\lambda_{s \ell} \gg \lambda_{b \ell}$ would
suggest an origin of flavor outside of symmetries, such as anarchy.
Corresponding final states from pair production are $jj \ell \ell^\prime$, where lepton species $\ell, \ell^\prime$ can be the same or different.

 Our analysis shows that the LHC, even with 3$\mbox{ab}^{-1}$ is not able to cover the full targeted parameter space, see figure \ref{fig:quark_gen_comparison}, where also expectations for future  $pp$-colliders with higher 
 center of mass-energy and comparable luminosity are shown.
 For measurements to be useful for flavor $b$-identification is required.  
Additionally, ability to  tag the flavor of the $b$ would allow to measure the leptoquark's electric charge and distinguish leptoquark representations.

\bigskip

{\bf Note added:} During the finalization of this work a related study on  leptoquark production at colliders appeared \cite{Dorsner:2018ynv}.

\section{Acknowledgements}
We are happy to thank Ulrik Egede, Yossi Nir  and Jos\' e Zurita for useful discussions. 
This project is supported in part 
by the {Bundesministerium f\"ur Bildung und Forschung (BMBF)}.

\end{document}